\documentclass[12pt]{article}

\usepackage[utf8]{inputenc}
\usepackage[T1]{fontenc}
\usepackage{times}
\usepackage{microtype}

\usepackage{subcaption}

\usepackage[margin=1in]{geometry}
\usepackage{setspace}
\usepackage{authblk}

\usepackage{amsmath}
\usepackage{amssymb}

\usepackage{booktabs}
\usepackage{array}
\usepackage{graphicx}
\usepackage{caption}
\usepackage{float}

\usepackage[round,authoryear,sort]{natbib}

\usepackage{hyperref}
\hypersetup{
    colorlinks = true,
    linkcolor  = black,
    citecolor  = blue,
    urlcolor   = blue,
    pdfauthor  = {Moses Boudourides},
    pdftitle   = {Structural Hallucination in Large Language Models: 
    A Network-Based Evaluation of Knowledge Organization and Citation Integrity}
}

\title{Structural Hallucination in Large Language Models:\\ A Network-Based Evaluation of Knowledge Organization and Citation Integrity}

\author{Moses Boudourides}
\affil{School of Professional Studies, Northwestern University\\
\small\url{Moses.Boudourides@northwestern.edu}}

\date{}

\begin{document}

\maketitle

\begin{abstract}
Large Language Models (LLMs) increasingly mediate access to scholarly information, yet their outputs are typically evaluated at the level of individual statements rather than knowledge structure. This paper introduces \emph{structural hallucination}: distortion of conceptual organization, relational architecture, and bibliographic grounding that remains invisible to sentence-level accuracy metrics. We develop a network-based stress test grounded in knowledge graph extraction, graph similarity analysis, centrality comparison, and citation integrity verification. The protocol is applied to three structured domains representing core forms of scholarly knowledge: Roget’s \emph{Thesaurus} (1911) as a historical knowledge organization system; Wikidata philosophers as a biographical knowledge graph; and bibliographic citation records from Dimensions.ai. Across all domains, substantial structural divergence is observed. In the lexical benchmark, pooled F1 scores range from 0.07 to 0.15; in the biographical benchmark, 40--90\% of generated values are absent from Wikidata and DBpedia; and in the bibliometric benchmark, citation omission reaches 100\%. Network-level comparison in the Roget reconstruction reveals node-set Jaccard similarity of 0.088 and a source-mismatch rate of 71\%. These findings show that structural fidelity cannot be inferred from local fluency. The proposed stress test provides a reproducible instrument for evaluating the structural integrity of LLM-generated knowledge representations within knowledge organization and information quality research.
\end{abstract}


\medskip
\section{Introduction}
\label{sec:introduction}

Large language models (LLMs) increasingly mediate access to scholarly information, but fluent output need not preserve the concepts, relations, or bibliographic links that organize a domain. This study examines that gap through \emph{structural hallucination}: relational mismatch that is not adequately captured by sentence-level accuracy measures. We develop a network-based stress test that represents selected reference sources and LLM outputs as comparable graphs, aligns them under benchmark-specific rules, and assesses node--edge overlap, network diagnostics, and citation integrity.

The framework is applied to the 1911 edition of Roget's \emph{Thesaurus}, Wikidata philosopher records, and Dimensions citation records. It asks whether the specified model and procedures recover the selected source structures, not whether all LLMs or knowledge organization systems behave alike. The contributions are a source-bounded account of structural hallucination, a reproducible comparison protocol, and evidence from three distinct benchmarks.

\section{The Epistemic Problem of Fluency, Error, and Accuracy}
\label{sec:fluency}

LLMs produce plausible continuations rather than verified retrievals \citep{Bender2021, Bommasani2021}. Their output can therefore be grammatical and contextually coherent while containing factual, bibliographic, or relational error. Hallucination taxonomies commonly distinguish intrinsic and extrinsic proposition-level falsehood \citep{Ji2023, Huang2023, Tonmoy2024}; the present study addresses the additional question of whether generated material preserves an externally documented relational structure.

This distinction complements recent evaluation of LLM-generated literature reviews. Tang, Duan, and Cai \citep{TangDuanCai2025v4} evaluate reference generation, literature-summary writing, and review composition, including generated-reference hallucination, semantic coverage, and factual consistency against human-written counterparts. The present framework instead evaluates alignment between generated structured representations and designated reference sources, using graph and network diagnostics. Fact-checking remains necessary, but verifying isolated claims cannot establish whether a resulting representation retains the conceptual or bibliographic organization of a field.

\medskip
\section{Structural Hallucination: Definition and Mechanisms}
\label{sec:structural}

Structural hallucination is a mismatch between the relational architecture represented in generated material and that represented in a documented comparison source. A generated statement may be locally plausible or correct while the combination of entities, conceptual dependencies, intellectual genealogies, citation links, or relation directions fails to recover the selected source structure. The term therefore concerns configuration, not merely proposition-level falsehood \citep{Maynez2020, Ji2023, Rawte2023}.

Three related manifestations motivate the analysis: conceptual re-centering, in which peripheral concepts are assigned undue structural prominence; bibliographic distortion, including source-nonmatched, omitted, or misattributed citations; and logical mis-structuring, in which locally plausible claims imply unsupported relations. These mechanisms can coexist and should be evaluated against an explicit source, graph-construction procedure, and matching rule. The following sections specify that evaluation design.

\medskip
\section{Evaluation Framework: Governance, Knowledge Graphs, and Citation Integrity}
\label{sec:governance}

\subsection{Governance as Reproducible Validation}

Here, governance is used in a restricted methodological sense: a validation design that makes generated representations inspectable against documented reference structures. It combines graph extraction and direct node--edge comparison, network diagnostics, and citation verification. Direct comparison records shared, missing, and generated-only entities or relations; network measures then describe differences in centrality and community structure. Citation checks examine reference existence, metadata, and omission relative to the selected record. These components are complementary: they do not convert a reference source into absolute truth, but they make the scope and basis of comparison explicit.

The subsequent benchmarks apply this design to lexical, biographical, and bibliographic sources. Their purpose is to test structural alignment under stated prompts, extraction procedures, matching rules, and operational reference sources.

\subsection{Reference Structures, Extraction, and Comparison}
\label{sec:kggovernance}

A knowledge organization system (KOS) is a curated conceptual and organizational scheme whose scope, provenance, representational rules, and community use shape a domain; a knowledge graph (KG) is a graph-based representation of entities or concepts and typed relations \citep{Hjorland2003, Zeng2008, Hjorland2021}. The two categories are distinct but overlapping. A KOS may be implemented or exported as a KG, and a KG may incorporate a KOS, but neither term is automatically synonymous with the other. Here, the reference graph is a documented export of, or a transparent derivation from, the selected source; the LLM graph is an analytic representation extracted from generated text. The comparison therefore evaluates two documented graph representations, not interchangeable categories or an absolute truth claim.

The reference source determines what the analysis can establish. Its scope, version, curatorial choices, omissions, and biases must be reported. When no canonical graph exists, researchers may use a transparent expert reference, multiple documented references, or a consensus procedure and report how conclusions depend on that choice. Information quality (IQ) is considered here at the relational level: accuracy, completeness, and consistency of graph structure rather than isolated attributes \citep{Stvilia2007, Arazy2011}.

\subsection{Matching and Network Diagnostics}
\label{sec:matching}

The analysis first extracts entities and typed relations from the generated material and the selected reference source, then aligns records under a benchmark-specific key. Node equivalence is established before graph intersection, difference, or centrality comparison. For generated node $u$ and reference node $v$ in benchmark $b$, let $r(u)$ and $r(v)$ denote the records in which the two values occur: a Roget Head, a philosopher, or a publication. The record-alignment indicator $A_b(r(u),r(v))$ equals 1 when the two records share the benchmark key, which is the normalised Head name in Roget, the exact philosopher label in Wikidata, and the publication title in Dimensions. The role function $\rho(u)$ returns the field or relation role that carries the value (for example \texttt{noun\_list}, \texttt{place\_of\_birth}, or \texttt{cited\_by}), so the condition $\rho(u)=\rho(v)$ restricts comparison to values of the same field. The value-equivalence test $E_b(u,v)$ is benchmark-specific. Writing $\nu(\cdot)$ for lower-casing and trimming, and $T(\cdot)$ for the set of tokens obtained by splitting $\nu(\cdot)$ at spaces, hyphens, and commas, the three tests are
\begin{align*}
E_{\mathrm{Roget}}(u,v)&=\mathbf{1}\big[\nu(u)=\nu(v)\big],\\
E_{\mathrm{Wiki}}(u,v)&=\mathbf{1}\Big[\nu(u)\sqsubseteq\nu(v)\ \lor\ \nu(v)\sqsubseteq\nu(u)\ \lor\ \frac{|T(u)\cap T(v)|}{\max(|T(u)|,|T(v)|)}\ge 0.6\Big],\\
E_{\mathrm{Dim}}(u,v)&=\mathbf{1}\Big[\frac{|T(u)\cap T(v)|}{|T(u)\cup T(v)|}\ge 0.6\Big],
\end{align*}
where $\sqsubseteq$ denotes substring containment. In Roget, each list entry is first split at commas and semicolons, and bracketed glosses, ``\&c'' markers, and numerals are removed, so that $u$ and $v$ are single terms or phrases rather than whole list segments; generated cross-references are stripped of a leading ``Head $n$.'' prefix. The matching rule is
\[
M_b(u,v)=1 \iff A_b(r(u),r(v))=1\;\land\;\rho(u)=\rho(v)\;\land\;E_b(u,v)=1.
\]
For list-valued fields, a generated value $u$ is a true positive if $M_b(u,v)=1$ for at least one reference value $v$ in the aligned record and field, and a false positive otherwise; a reference value with no matching generated value is a false negative. This strict operational matching is deliberate: it makes the comparison reproducible and auditable, prevents an unreported similarity model from silently converting non-identical values into matches, and tests whether the generated output recovers the designated reference structure under a stated schema. Candidate-generation aids such as semantic similarity may support review but do not by themselves establish a match.

The resulting node--edge comparison identifies shared, missing, and generated-only elements. Network diagnostics then compare centrality, community structure, and graph similarity among the documented graphs. This order separates what is present or absent under the matching rule from how retained and generated structures are organized.

\subsection{Running Example: Kuno Fischer}
\label{sec:kuno-fischer-running-example}

For Kuno Fischer, the cached Wikidata reference record gives birth date \texttt{1824-07-23}; the stored reconstruction gives \texttt{1824-04-22}. Under the local date rule, the values share only the year token and do not match. This illustrates a source-derived and a generated attribute edge that diverge before any graph-level measure is calculated; it does not itself classify the generated value as a hallucination.

Applying the rule term by term, let $u=\texttt{1824-04-22}$ and $v=\texttt{1824-07-23}$. Both values occur in the record of the same philosopher, so $r(u)$ and $r(v)$ carry the identical label ``Kuno Fischer'' and $A_{\mathrm{Wiki}}(r(u),r(v))=1$. Both are values of the field \texttt{birth}, so $\rho(u)=\rho(v)$. For $E_{\mathrm{Wiki}}$, neither string contains the other; tokenisation gives $T(u)=\{1824,04,22\}$ and $T(v)=\{1824,07,23\}$, so $|T(u)\cap T(v)|/\max(|T(u)|,|T(v)|)=1/3<0.6$ and $E_{\mathrm{Wiki}}(u,v)=0$. Hence $M_{\mathrm{Wiki}}(u,v)=0$: the reference value is counted as a false negative and the generated value as a false positive. The generated value is then looked up among the DBpedia \texttt{birthDate} values for the same resource; only if that lookup also fails is it counted as a hallucination in the sense used in Table~\ref{tab:phil_classical}. The same record shows the case $A_{\mathrm{Wiki}}=1$ and $\rho(u)=\rho(v)$ with an empty reference value: the reference \emph{influenced\_by} field is empty while the reconstruction supplies \emph{Hegel} and \emph{Kant}, so no reference value $v$ exists and both generated values are false positives.

The implementation checklist, full notation key, pseudocode, fixed prompts, and matching-threshold sensitivity analyses are supplied in Supplementary Appendix S1, which is appended to this manuscript.

\subsection{Citation Integrity and Bibliometric Grounding}
\label{sec:citation}

Citation integrity is the bibliographic component of structural evaluation. Generated references should be checked against authoritative bibliographic sources for existence, metadata, and the relevance of the cited work to the claim it supports \citep{Alkaissi2023, OrdunaMailea2023}. At the structural level, the generated citation graph can be compared with a documented reference graph for missing works, altered citation neighborhoods, centrality differences, and uneven author or venue representation. These analyses assess relational alignment rather than local correctness alone. In the Dimensions benchmark below, this comparison is applied to the citing-record neighbourhood of each publication: a generated citing record counts as recovered only if it matches a reference citing record by DOI or by title under the Dimensions equivalence test of Section~\ref{sec:matching}.

\medskip
\section{Empirical Benchmarks: Three Stress Tests of Structural Hallucination}
\label{sec:benchmarks}

\subsection*{Shared Evaluation Pipeline}

The three benchmarks use source-specific operational reference baselines, not exhaustive accounts of all valid knowledge. For each selected reference record, \texttt{gpt-4.1-mini} was prompted to reconstruct the relevant structured fields in JSON. Responses were parsed into the reference schema, cached after the first successful call, and retried with exponential backoff when necessary. The evaluated model is the research object; the manuscript's revision-assistance disclosure appears in the end matter.

Records were aligned by normalized Head name (Roget), philosopher name (Wikidata), or publication title (Dimensions). Alignment does not establish value equivalence. Values were compared only within the aligned record, field, and relation role. The implementation does not perform one-to-one assignment for list-valued fields: a value is classified by whether at least one eligible match exists. Multiple eligible values require a documented tie-break or review before one-to-one structural comparison.

Roget uses lower-cased, stripped exact membership of single terms. Wikidata uses lower-cased, stripped values and accepts containment or shared-token proportion at least 0.60; a local nonmatch is checked against DBpedia, yielding either verified extra knowledge or a value absent from both sources. Dimensions uses lower-cased, trimmed token-set Jaccard similarity of at least 0.60, while DOI resolution, citation-title similarity, year match, and DOI validity remain separate diagnostics. Precision, recall, F1, and accuracy follow these benchmark-specific procedures.

All three benchmarks use the same aggregation. For each field, true positives (TP), false positives (FP), false negatives (FN), and, where both values are empty, true negatives (TN) are summed over all aligned records before any ratio is formed, so every reported precision $P=\mathrm{TP}/(\mathrm{TP}+\mathrm{FP})$, recall $R=\mathrm{TP}/(\mathrm{TP}+\mathrm{FN})$, and $F_1=2PR/(P+R)$ is a pooled (micro-averaged) score for that field; no per-record scores are averaged. Accuracy is $(\mathrm{TP}+\mathrm{TN})/(\mathrm{TP}+\mathrm{FP}+\mathrm{FN}+\mathrm{TN})$; for the Roget term lists, which are compared term by term and therefore have no true-negative category, it reduces to $\mathrm{TP}/(\mathrm{TP}+\mathrm{FP}+\mathrm{FN})$. In the Wikidata benchmark, the false positives are further split into values confirmed by DBpedia ($\mathrm{FP}_{\mathrm{extra}}$) and values absent from both sources ($\mathrm{FP}_{\mathrm{hall}}$), and the hallucination rate is $\mathrm{FP}_{\mathrm{hall}}/(\mathrm{TP}+\mathrm{FP})$, the share of all generated values that neither source supports. The 95\% confidence intervals for $F_1$ are percentile intervals from 2,000 bootstrap resamples of records (Heads, philosophers, or publications) \citep{EfronTibshirani1993}; intervals for the hallucination rate are Wilson score intervals \citep{Wilson1927}. Matching-threshold sensitivity analyses for Wikidata and Dimensions (thresholds from 0.4 to 1.0) and a normalisation-sensitivity analysis for Roget are reported in Supplementary Appendix S1; the conclusions stated below hold across these variants except where noted.

Semantic similarity uses \texttt{all-MiniLM-L6-v2} as complementary contextual evidence, not as an equivalence decision. An exploratory logistic-regression classifier used in earlier versions of this analysis is not reported, because its training label was defined from the same similarity and missing-value indicators that served as its predictors, so its discrimination could not be informative.

\subsection{Benchmark 1: Lexical Reconstruction of Roget's Thesaurus}
\label{sec:roget_results}

The first benchmark uses the 1911 edition of Roget's \emph{Thesaurus of
English Words and Phrases} \citep{Roget1911}, obtained from Project
Gutenberg. Roget's \emph{Thesaurus} is a historically situated,
hierarchically organized lexical knowledge organization system (thesaurus)
that arranges English words and phrases into Classes, Sections, and Heads.
Each Head functions as a lexical-semantic grouping and is associated with
curated lists of nouns, verbs, adjectives, adverbs, and cross-references to
related Heads. The 1911 edition contains 1,000 Heads.

\paragraph{Source extraction, sampling, and generation.}
We treated the 1911 edition as a historical reference source rather than an unqualified reference baseline. A custom Python parser processed Project Gutenberg transcription No.~10681, identified numbered Head boundaries, excluded sub-entries and malformed or empty records, and returned 997 usable Heads. It extracted noun, verb, adjective, and adverb sections as semicolon-split, trimmed list strings; cross-references came from \texttt{\&c <number>} markers and were mapped to Head names where possible.

With Python seed 42, we sampled 30 Heads without replacement. The routine obtained one \texttt{gpt-4.1-mini} JSON response per Head (or reused a cached response), with temperature 0 and a 1,024-token maximum. The archived run has 30 successful records but no request timestamp or full provider metadata. The prompt requested \texttt{class}, \texttt{section}, \texttt{head\_name}, the four part-of-speech lists, and \texttt{cross\_references}, and asked that each list be kept to 10--20 items; the fixed prompt is reproduced in Supplementary Appendix S1. Because a sampled Head contains on average 92 reference terms, this length instruction bounds the recall that any response could reach: if every generated term were correct, pooled recall could not exceed 0.40 (nouns), 0.54 (verbs), 0.59 (adjectives), or 0.54 (adverbs). Precision is not affected by this bound.

Reference and generated records were aligned by normalized \texttt{head\_name}. For each aligned field, the semicolon-split list strings of both records were further split into single terms as described in Section~\ref{sec:matching}, and the lower-cased, stripped terms were compared by exact set membership; missing or non-list values were empty sets. The same term sets define the nodes of the knowledge graphs analysed below, so the field-level and network-level results rest on one normalisation. Precision, recall, and F1 pool TP, FP, and FN across the 30 Heads, so they are micro exact-string scores by field. Synonymy, inflection, aliases, and semantic similarity did not establish matches; generated-only and reference-only strings are therefore protocol-specific mismatch categories, not factual-falsehood verdicts.

\subsubsection{Classical Evaluation}

Classical evaluation, using exact membership in normalised term sets, indicates low reconstruction fidelity across all fields, with pooled F1 between 0.07 and 0.15 (Table~\ref{tab:roget_classical}).

\begin{table}[H]
\centering
\caption{Classical evaluation results for Roget's Thesaurus reconstruction
($n=30$ Heads). TP, FP, and FN are summed over Heads before the pooled (micro-averaged) scores are formed; accuracy is $\mathrm{TP}/(\mathrm{TP}+\mathrm{FP}+\mathrm{FN})$. The F1 interval is a 95\% bootstrap interval over Heads.}
\label{tab:roget_classical}
\small
\begin{tabular}{lrrrccccc}
\toprule
\textbf{Field} & \textbf{TP} & \textbf{FP} & \textbf{FN} & \textbf{Precision} & \textbf{Recall} & \textbf{F1} & \textbf{95\% CI (F1)} & \textbf{Accuracy} \\
\midrule
noun\_list        & 121 & 254 & 1,077 & 0.323 & 0.101 & 0.154 & [0.117, 0.195] & 0.083 \\
verb\_list        &  53 & 207 &   561 & 0.204 & 0.086 & 0.121 & [0.066, 0.171] & 0.065 \\
adjective\_list   &  74 & 212 &   604 & 0.259 & 0.109 & 0.154 & [0.118, 0.190] & 0.083 \\
adverb\_list      &  15 & 139 &   253 & 0.097 & 0.056 & 0.071 & [0.025, 0.137] & 0.037 \\
cross\_references &   8 &  96 &    68 & 0.077 & 0.105 & 0.089 & [0.032, 0.159] & 0.047 \\
\bottomrule
\end{tabular}
\end{table}

\begin{figure}[H]
\centering
\includegraphics[width=0.9\textwidth]{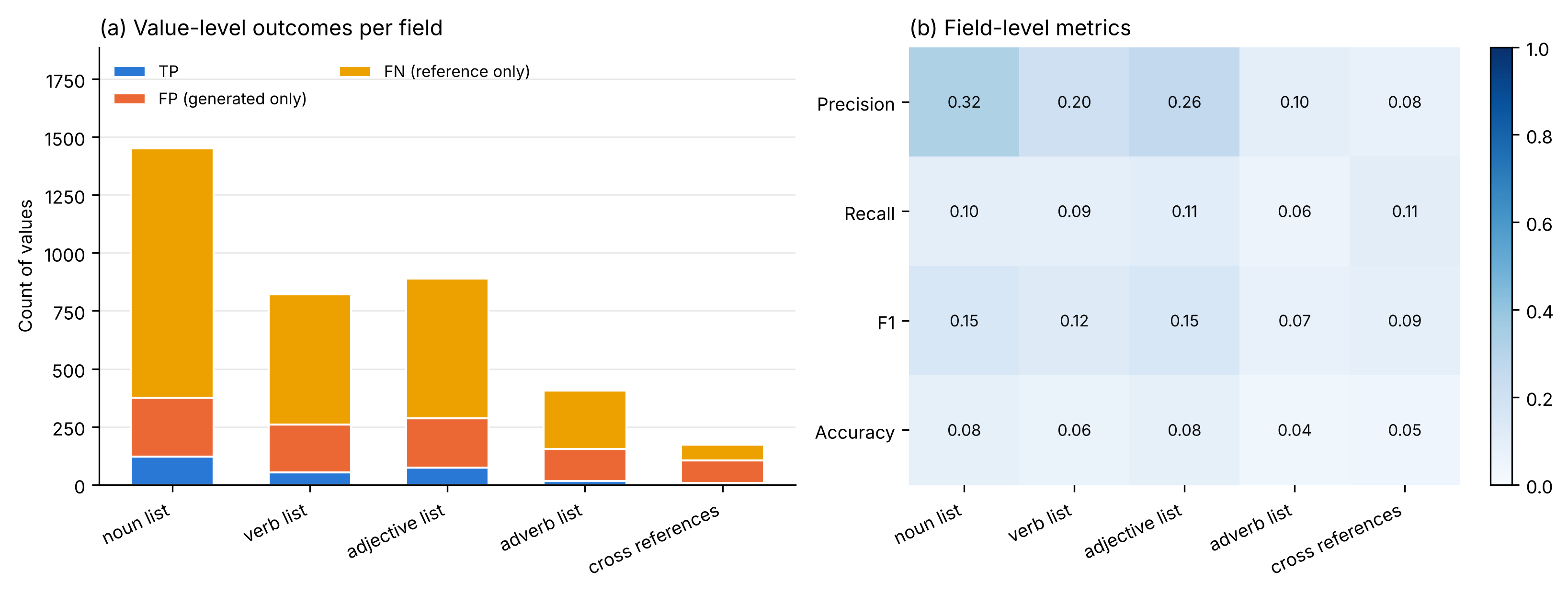}

\smallskip
\caption{Classical evaluation of LLM reconstruction of Roget's Thesaurus ($n = 30$ Heads). Panel (a) shows the pooled TP, FP (generated only), and FN (reference only) counts per field; panel (b) shows the pooled precision, recall, F1, and accuracy reported in Table~\ref{tab:roget_classical}. False negatives dominate every term field, partly because the prompt limited each generated list to 10--20 items (Section~\ref{sec:roget_results}).}
\label{fig:roget_classical}
\end{figure}

Across the five fields, pooled exact-term F1 ranges from 0.071 (adverbs) to 0.154 (nouns and adjectives). Recall (0.056--0.109) lies far below the ceiling imposed by the list-length instruction (0.40--0.59), and precision, which that instruction does not constrain, remains between 0.077 and 0.323: at most about one generated term in three appears in the corresponding 1911 Head. The result is a low recovery of the terms and links recorded in the 1911 edition under the specified exact matching protocol. It does not determine whether every generated-only term is semantically unrelated or factually false, and historical lexical change is a possible contributor to mismatch.

\paragraph{Representative cross-reference cases.} For \emph{Greatness} (Head 31), the reference lists \emph{Infinity}, \emph{Importance}, \emph{Superiority}, \emph{Multitude}, \emph{Completeness}, and \emph{Wonder}; the cached reconstruction returns \emph{Power}, \emph{Influence}, \emph{Eminence}, and \emph{Excellence}, yielding zero exact matches. For \emph{Permission} (Head 760), the reference relation is \emph{Consent}; the model returns \emph{Law}, \emph{Authority}, \emph{Power}, and \emph{Consent}, yielding one exact match and three model-only values. These are mismatches from the structure recorded in the 1911 edition, not proof that every generated relation is false.

Supplementary Figure S1 reports the Roget list-level semantic-similarity analysis. It is complementary descriptive evidence and was not used to establish exact matches or change the reported exact-term counts. Supplementary Table~\ref{tab:s_norm} shows that removing the obsolescence daggers and trailing punctuation that the 1911 transcription attaches to some terms raises the pooled F1 by at most 0.012, so the scores in Table~\ref{tab:roget_classical} are not an artefact of transcription markup.

\subsubsection{Network-Analytic Hallucination Stress Test}

\paragraph{Graph Structure and Source-Mismatch Rate.}
The directed knowledge graphs constructed from the Roget and LLM data exhibit
dramatically different structural properties (Table~\ref{tab:kg_summary}).

\begin{table}[H]
\centering
\caption{Structural comparison of the Roget and LLM directed knowledge graphs. Both graphs are built from the same single-term normalisation used in Table~\ref{tab:roget_classical}. Louvain communities are computed on the undirected projection with a fixed random seed.}
\label{tab:kg_summary}
\begin{tabular}{lcc}
\toprule
\textbf{Metric} & \textbf{Roget Graph} & \textbf{LLM Graph} \\
\midrule
Nodes                         & 2,708 & 1,066 \\
Edges                         & 2,760 & 1,093 \\
Communities (Louvain)         & 28    & 28    \\
Modularity $Q$                & 0.932 & 0.944 \\
Weakly connected components   & 6     & 18    \\
Mean term nodes per sampled Head & 92.0 & 36.4 \\
Cross-reference edges         & 76    & 104   \\
Term nodes linked to $\ge 2$ Heads & 47 & 26 \\
\midrule
Shared nodes / node Jaccard similarity $J$   & \multicolumn{2}{c}{305 / 0.088} \\
Shared edges / edge Jaccard similarity $J$   & \multicolumn{2}{c}{270 / 0.075} \\
LLM-only nodes (source-nonmatched) & \multicolumn{2}{c}{761 (71.4\,\%)} \\
\bottomrule
\end{tabular}
\end{table}

All counts in Table~\ref{tab:kg_summary} are computed over the same node sets: with 2,708 Roget nodes, 1,066 LLM nodes, and 305 shared nodes, the union contains 3,469 nodes, so $J=305/3{,}469=0.088$, and $1{,}066-305=761$ LLM nodes (71.4\,\%) have no counterpart in the Roget graph. Only 8.8\,\% of the nodes of the two graphs are shared.

This low overlap is not the overlap one would expect from two unrelated word lists of these sizes. For each sampled Head, the Jaccard similarity between its generated and its reference term set has a mean of 0.081 (median 0.080, range 0 to 0.195; Supplementary Table~\ref{tab:s_heads}); when each generated Head is instead compared with the reference terms of the other 29 Heads, the mean falls to 0.0005. A paired Wilcoxon signed-rank test ($p=1.3\times10^{-6}$) and a permutation test that randomly reassigns generated Heads to reference Heads (10,000 permutations, $p<10^{-4}$) both reject the hypothesis that the generated terms are unrelated to the Head they were requested for. The reconstruction is therefore on-topic for every Head but one, yet it recovers only a small part of each Head's documented vocabulary.

\begin{figure}[H]
\centering
\begin{subfigure}[t]{0.48\textwidth}
  \includegraphics[width=\textwidth]{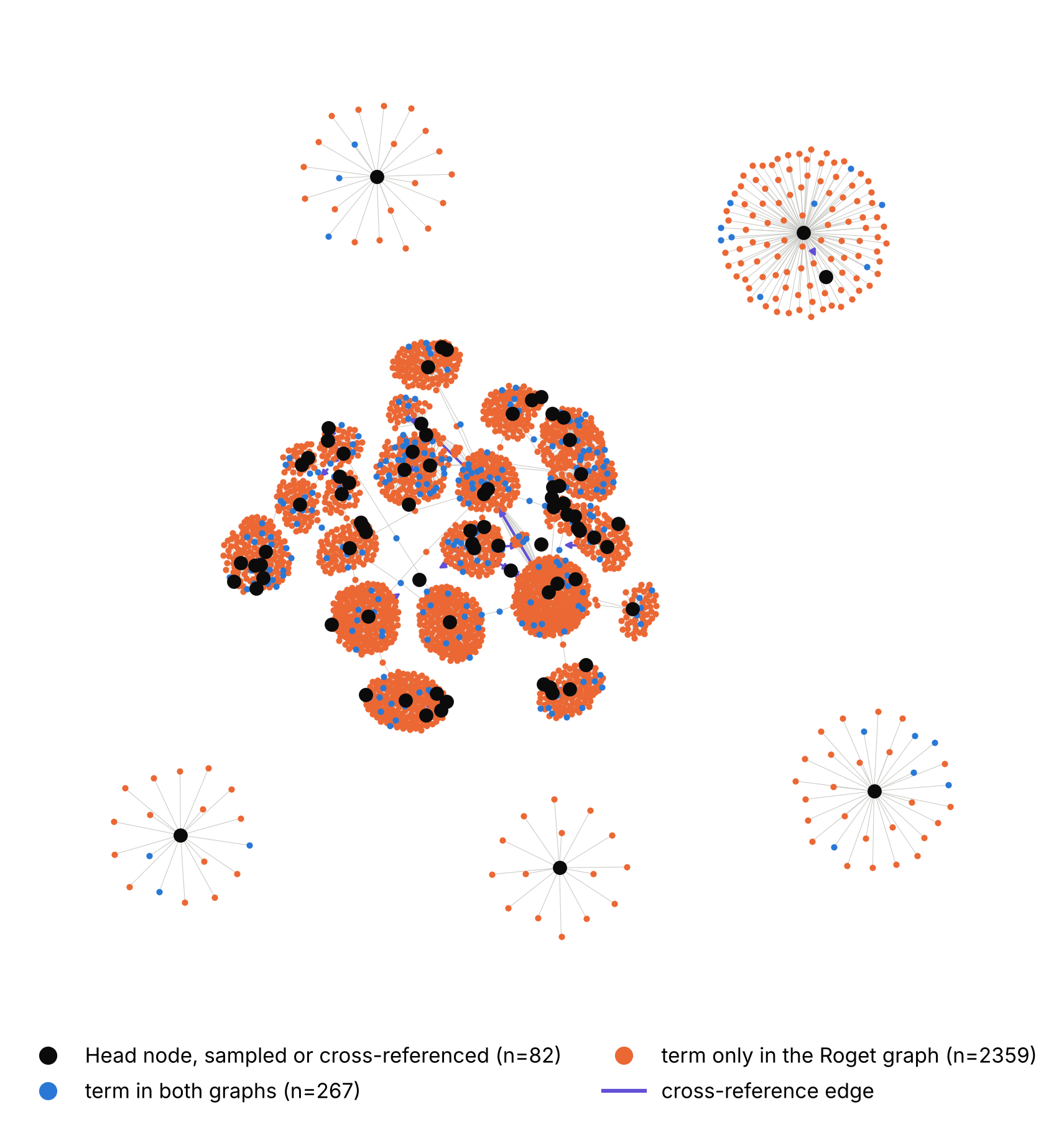}
  
\smallskip
\caption{Roget directed knowledge graph. Black nodes are Head nodes (sampled
  or cross-referenced); blue nodes are terms that also occur in the LLM graph;
  orange nodes are terms that occur only in the Roget graph. Grey edges link
  a Head to its terms; purple edges are cross-references between Heads.}
  \label{fig:roget_kg}
\end{subfigure}
\hfill
\begin{subfigure}[t]{0.48\textwidth}
  \includegraphics[width=\textwidth]{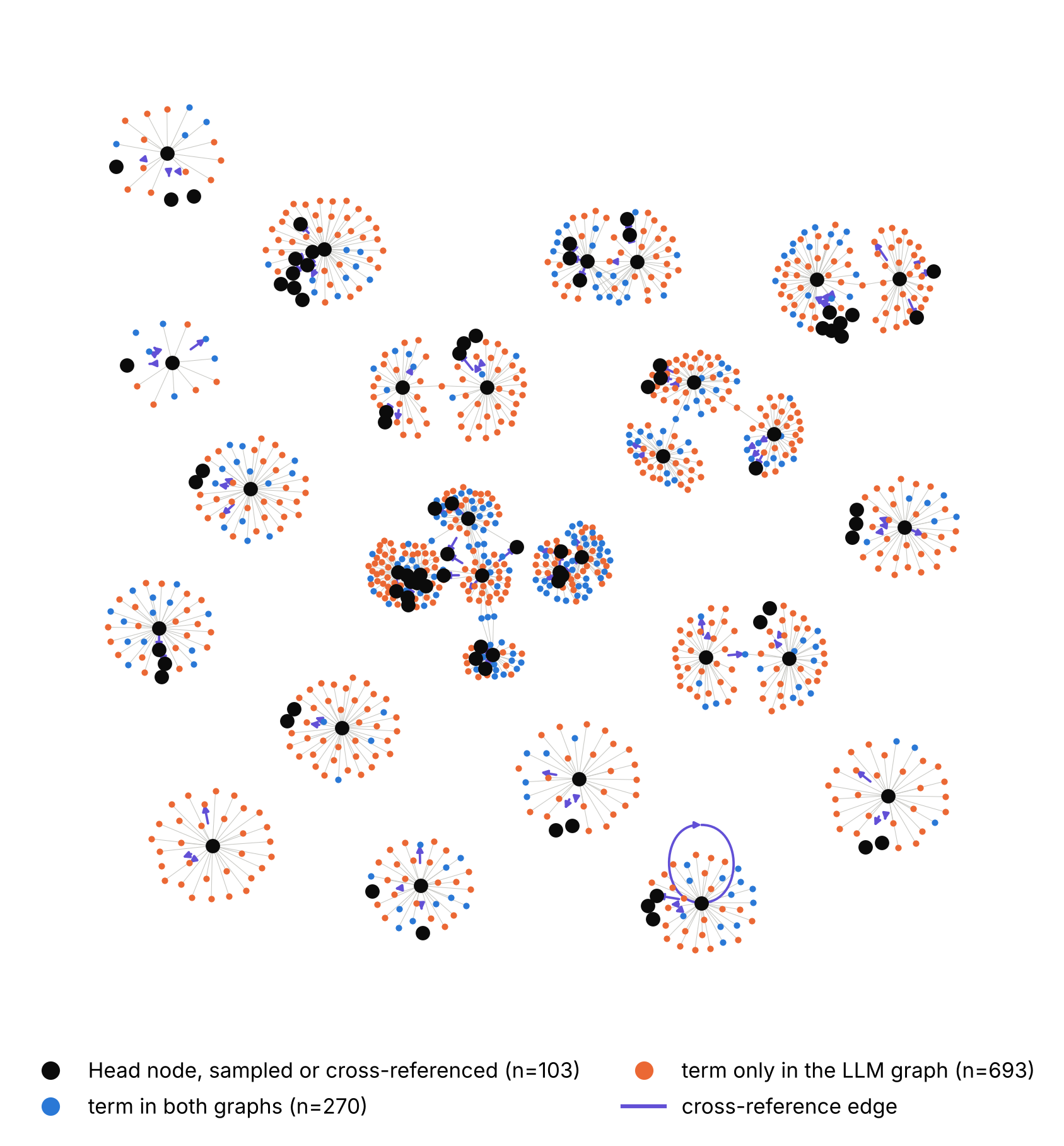}
  
\smallskip
\caption{LLM-generated directed knowledge graph, with the same colour scheme:
  blue terms also occur in the Roget graph, and orange terms occur only in the
  LLM graph and are therefore source-nonmatched.}
  \label{fig:llm_kg}
\end{subfigure}

\smallskip
\caption{Directed knowledge graphs for the Roget (left) and LLM-generated (right)
datasets, drawn with the graphviz \texttt{sfdp} layout. Each star-shaped cluster
represents one of the 30 sampled Heads. The star shape follows from the
graph-construction rule (every term is attached to the Head that lists it);
the figure is informative about which terms are shared, not about the
model's choice of topology.}
\label{fig:kg_networks}
\end{figure}

The visual comparison of the two graphs (Figure~\ref{fig:kg_networks}) shows
graphs of similar shape (star-shaped communities connected by cross-reference
edges, modularity $Q = 0.944$ vs.\ $0.932$) but different content (71.4\,\%
source-nonmatched nodes). Much of this similarity in shape is guaranteed by the
experimental representation rather than produced by the model. The prompt fixes
one record per Head with four term lists and a cross-reference list, and the
construction rule turns each record into a star whose centre is the Head;
any comparably prompted output would therefore yield one high-modularity star
per Head. What the construction leaves free to vary, and what can be attributed
to the model, is the content and size of each star and the links between
stars. On these quantities the two graphs differ: the generated Heads carry on
average 36.4 rather than 92.0 terms, the generated graph falls into 18 rather
than 6 weakly connected components, and it contains more cross-reference edges
(104 vs.\ 76) but fewer terms shared between Heads (26 vs.\ 47). The reading
that the model reproduces the \emph{form} of a thesaurus while failing to
reproduce its \emph{substance} is therefore best treated as a hypothesis about
these free quantities, not as a finding established by the similar topology.

\paragraph{Centrality Analysis.}
Centrality comparisons in these graphs must be read against the construction
rule. Every edge leaves a Head, so a term node has no out-edges, and PageRank
divides a Head's score equally among the terms it lists. The PageRank of a term
listed under a single Head is therefore fixed by the number of terms under that
Head: across such terms, the Spearman correlation between PageRank and the
parent Head's out-degree is $-1.00$ in the Roget graph and $-0.998$ in the LLM
graph. The highest-ranked terms are simply the terms of the shortest lists. In
the Roget graph they are the 15 terms of \emph{Dextrality}; in the LLM graph
they are the 12 terms generated for \emph{Brown}, most of which (\emph{brown},
\emph{tawny}, \emph{chestnut}, \emph{cinnamon}, \emph{auburn}, \emph{russet},
\emph{sepia}) also occur in Roget's own \emph{Brown} Head. A comparison of
top-ranked PageRank nodes across the two graphs thus pairs terms from
different Heads and carries no information about conceptual re-centering; we
do not interpret it. Over the 305 shared nodes, PageRank values in the two
graphs are only weakly correlated (Spearman $\rho=0.14$, $p=0.015$).

Betweenness is equally constrained. A directed path of length two or more
exists only where a cross-reference leads from one sampled Head to another
sampled Head, so betweenness is non-zero for only four Roget nodes and three
LLM nodes; all other nodes are tied at zero. Table~\ref{tab:betweenness}
therefore reports, reciprocally, every node with non-zero betweenness in either
graph together with its value in the other graph. Three of the four Roget
intermediaries (\emph{Superiority}, \emph{Repetition}, \emph{Permission}) keep
that role in the generated graph, while \emph{Greatness} loses it; the generated
graph adds no intermediary of its own.

\begin{table}[H]
\centering
\caption{Reciprocal betweenness comparison: every node with non-zero
betweenness centrality in either directed graph, with its value in both
graphs (values $\times10^{-5}$). All other nodes have zero betweenness in both
graphs.}
\label{tab:betweenness}
\begin{tabular}{lcc}
\toprule
\textbf{Node (Head)} & \textbf{Roget graph} & \textbf{LLM graph} \\
\midrule
Superiority & 7.58 & 3.62 \\
Greatness   & 7.40 & 0.00 \\
Repetition  & 1.56 & 1.85 \\
Permission  & 1.41 & 3.79 \\
\bottomrule
\end{tabular}
\end{table}

\begin{figure}[H]
\centering
\includegraphics[width=\textwidth]{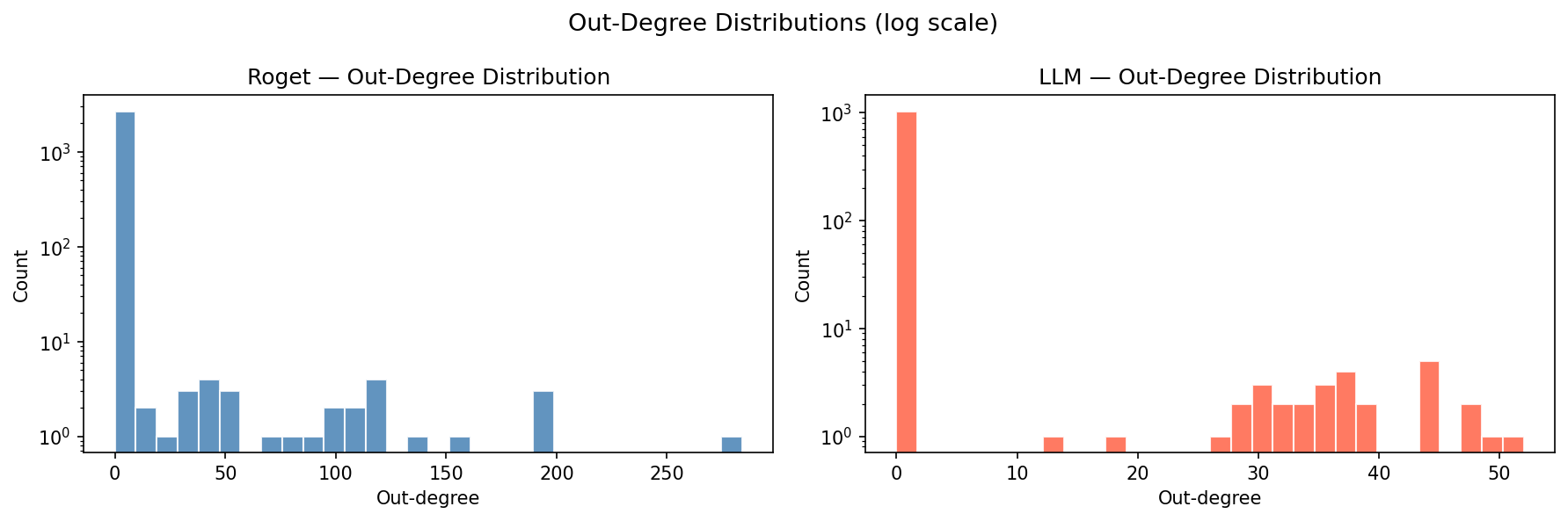}

\smallskip
\caption{Out-degree distributions for the Roget (left) and LLM-generated (right)
knowledge graphs, on a logarithmic count scale. In both graphs, term nodes have
out-degree zero by construction, and the remaining mass corresponds to Head
nodes, whose out-degree equals the number of terms and cross-references listed
for them. The generated Heads have lower and less dispersed out-degrees
(at most about 50, against more than 250 in the Roget graph), consistent with
the 10--20-item list instruction.}
\label{fig:kg_degree}
\end{figure}

The quantities on which the graphs genuinely differ are therefore list length,
term content, and the few inter-Head links, not the centrality hierarchy. For
applications that would use LLM-generated knowledge graphs for semantic search,
concept recommendation, or ontology alignment \citep{Paulheim2017, Ehrlinger2016},
the relevant risk shown here is the substitution and omission of terms within
an otherwise well-formed structure.

\subsubsection{Historical Lexical Fidelity and Temporal Bias}

The lexical divergence in Benchmark 1 is consistent with a failure of historical lexical fidelity. The 1911 Roget's \emph{Thesaurus} encodes a period-specific knowledge organization system, and the generated entries draw on present-day vocabulary. The clearest case is \emph{Dextrality}, the only sampled Head with no shared term (Table~\ref{tab:lexical_divergence_examples}). Roget uses the Head for right-handedness and the right side (\emph{right hand}, \emph{dexter}, \emph{dextral}, \emph{starboard}, \emph{offside}), whereas the model returns the vocabulary of manual skill (\emph{dexterity}, \emph{adroitness}, \emph{deftness}, \emph{nimble}, \emph{handy}), a sense that the word does not carry in the 1911 classification. For a concrete physical Head such as \emph{Brown}, by contrast, seven of the twelve generated terms occur in the reference list.

\begin{table}[H]
\centering
\caption{Within-Head comparison of reference and generated terms for two sampled Heads. Terms are shown after the normalisation of Section~\ref{sec:matching}; the dagger is the 1911 edition's mark for words already obsolete at the time.}
\label{tab:lexical_divergence_examples}
\small
\begin{tabular}{>{\raggedright\arraybackslash}p{0.12\textwidth}>{\raggedright\arraybackslash}p{0.35\textwidth}>{\raggedright\arraybackslash}p{0.35\textwidth}c}
\toprule
\textbf{Head} & \textbf{Roget 1911 terms (selection)} & \textbf{Generated terms (selection)} & \textbf{$J$} \\
\midrule
Dextrality & right hand, dexter, dextral, right-handed, starboard, offside, dextrorsal$^\dagger$, dextrorse$^\dagger$, ambidextrous & dexterity, dexterous, adroitness, deftness, skill, nimble, handy, finesse, manual dexterity & 0.000 \\
\addlinespace
Brown & brown, auburn, chestnut, cinnamon, russet, sepia, tawny, bay, maroon, chocolate, nut-brown, fuscous$^\dagger$ & brown, auburn, chestnut, cinnamon, russet, sepia, tawny, hazel, mahogany, umber, brownish & 0.189 \\
\bottomrule
\end{tabular}
\end{table}

Two tests make the temporal reading quantitative. First, the 1911 edition marks with a dagger the words that were already obsolete when it was published \citep{RogetGutenberg22}. Of the 2,758 reference terms in the four term fields, 262 carry this mark; none of them is recovered, whereas 269 of the 2,496 unmarked terms (10.8\,\%) are (Fisher's exact test, $p=1.0\times10^{-12}$). Second, using Zipf-scale frequencies of present-day English from the \texttt{wordfreq} library \citep{Speer2022}, the reference terms that the model misses are less frequent today than the reference terms it recovers (median Zipf 3.35 vs.\ 3.56; one-sided Mann--Whitney test, $p=2\times10^{-4}$), and less frequent than the terms that the model adds (median 3.58, $p=1.4\times10^{-5}$). The frequency differences are modest in size, but their direction, together with the complete absence of obsolete terms, indicates that what the model retains from a historical vocabulary is biased toward words still current.

These results describe the output; they do not identify a mechanism. A plausible conjecture is that the substitutions reflect the distribution of present-day usage in the training data, so that the model interpolates toward contemporary lexical regularities and produces an anachronistic reconstruction that preserves structural form while overwriting historical content, an instance of ``ontological drift'' \citep{Tennis2002}. The design contains no manipulation of training data or of temporal distribution, so this explanation remains a hypothesis to be tested, for instance with models of known training cut-offs or with period-specific corpora.

This pattern can be further interpreted through Hjørland’s
domain-analytic approach, which holds that knowledge organization systems
must reflect the historically situated epistemic norms of specific
scholarly domains rather than abstract or universal semantic regularities
\citep{Hjorland2002, Hjorland2017}. From a domain-analytic perspective,
Roget’s 1911 knowledge organization system embodies the conceptual structure, classificatory
logic, and lexical conventions of its historical moment. The LLM’s
replacement of period-specific terms with statistically dominant
contemporary synonyms can therefore be read not merely as lexical substitution
but as a failure to respect domain-bound historical semantics. The resulting
“ontological drift” reflects a shift from domain-specific epistemic
organization to corpus-wide distributional generalisation. In this sense,
structural hallucination can be understood as a breakdown in
domain-sensitive representation: the model optimises for cross-domain
statistical plausibility rather than fidelity to the internal historical
logic of a particular knowledge organization system.

\subsection{Benchmark 2: Biographical Knowledge --- Wikidata Philosophers}
\label{sec:philosophers_results}

The second benchmark evaluates the LLM's capacity to reconstruct structured
biographical knowledge for 1,532 philosophers born between 1800 and 1850,
sourced from Wikidata \citep{Vrandecic2014}. After matching LLM responses to
Wikidata entries, 1,527 matched records were available for evaluation. The six
fields evaluated are: birth date, death date, place of birth, country of
citizenship, \emph{influenced\_by}, and field of work. The model received the
names in batches of 20 under the fixed prompt reproduced in Supplementary
Appendix S1.

The record-level comparison in Section~\ref{sec:kuno-fischer-running-example} illustrates how one source-to-output field difference is represented before the aggregate results below are calculated across the full matched set.

\subsubsection{Results}

\begin{table}[H]
\centering
\caption{Classical evaluation results for Wikidata philosopher reconstruction
($n=1{,}527$ matched records). All scores are pooled over records. Generated
values are $\mathrm{TP}+\mathrm{FP}$; the hallucination rate is
$\mathrm{FP}_{\mathrm{hall}}/(\mathrm{TP}+\mathrm{FP})$, the share of all
generated values that match neither the Wikidata record nor DBpedia. Accuracy is
$(\mathrm{TP}+\mathrm{TN})/(\mathrm{TP}+\mathrm{FP}+\mathrm{FN}+\mathrm{TN})$.
Intervals are 95\% bootstrap intervals over records (F1) and Wilson intervals
(hallucination rate).}
\label{tab:phil_classical}
\footnotesize
\setlength{\tabcolsep}{3.5pt}
\begin{tabular}{lcccccrc}
\toprule
\textbf{Field} & \textbf{P} & \textbf{R} & \textbf{F1 [95\% CI]} & \textbf{Acc.} & \textbf{Generated} & \textbf{FP$_{\mathrm{hall}}$} & \textbf{Hall.\ rate [95\% CI]} \\
\midrule
birth                    & 0.258 & 0.183 & 0.214 [0.193, 0.238] & 0.120 & 1,085 &   795 & 0.733 [0.706, 0.758] \\
death                    & 0.262 & 0.185 & 0.217 [0.195, 0.239] & 0.121 & 1,076 &   787 & 0.731 [0.704, 0.757] \\
place\_of\_birth         & 0.144 & 0.129 & 0.136 [0.118, 0.156] & 0.112 & 1,134 &   911 & 0.803 [0.779, 0.825] \\
country\_of\_citizenship & 0.585 & 0.588 & 0.586 [0.560, 0.612] & 0.452 & 1,204 &   487 & 0.404 [0.377, 0.432] \\
influenced\_by           & 0.092 & 0.117 & 0.103 [0.072, 0.138] & 0.641 &   424 &   380 & 0.896 [0.864, 0.922] \\
field\_of\_work          & 0.185 & 0.326 & 0.236 [0.216, 0.257] & 0.235 & 1,930 & 1,550 & 0.803 [0.785, 0.820] \\
\bottomrule
\end{tabular}
\end{table}

\begin{figure}[H]
\centering
\includegraphics[width=\textwidth]{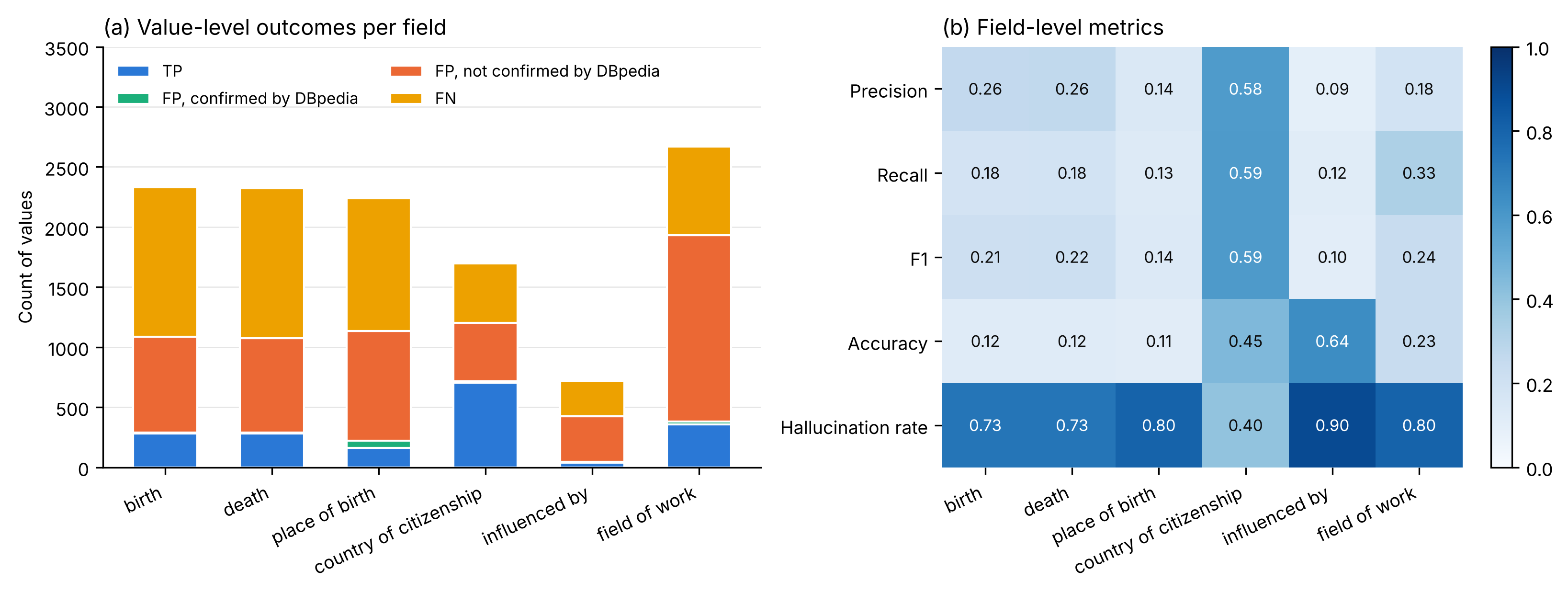}

\smallskip
\caption{Classical evaluation of LLM reconstruction of Wikidata philosopher
data ($n=1{,}527$ matched records). Panel (a) splits the generated values that
do not match the Wikidata record into those confirmed by DBpedia and those not
confirmed by either source; panel (b) shows the pooled precision, recall, F1,
accuracy, and hallucination rate of Table~\ref{tab:phil_classical}. Country of
citizenship achieves the highest F1 (0.586) and the lowest hallucination rate
(0.404); \emph{influenced\_by} achieves the lowest F1 (0.103) and the highest
hallucination rate (0.896). The high accuracy of \emph{influenced\_by} (0.641)
reflects the many records in which both sources leave the field empty.}
\label{fig:phil_classical}
\end{figure}

The results reveal a consistent pattern: the model performs better on
categorical fields with a small, well-defined value space (country of
citizenship: F1 = 0.586) and worse on set-valued fields requiring specific
factual recall (\emph{influenced\_by}: F1 = 0.103; place of birth: F1 = 0.136).
Between 40.4\,\% (country of citizenship) and 89.6\,\% (\emph{influenced\_by}) of the generated values in each field match neither the Wikidata record nor DBpedia. DBpedia confirms very few of the generated values that miss the Wikidata record (between 5 and 60 per field), so among these unmatched values the share left unconfirmed is 93.8\,\% to 99.1\,\%; this conditional share, not the share of all generated values, is what the first version of this analysis reported. The non-overlapping confidence intervals in Table~\ref{tab:phil_classical} show that the advantage of country of citizenship over every other field, in both F1 and hallucination rate, is not attributable to sampling variation across philosophers.

The death date result is particularly striking. The model achieves an F1 of only
0.217 for death dates, despite this being a scalar field with a single correct
answer. The hallucination rate of 73.1\,\% means that nearly three in four death dates
provided by the model are absent from both Wikidata and DBpedia, and DBpedia
confirms only 7 of the 794 generated death dates that miss the Wikidata record.
The low F1 and high hallucination rate are consistent with generative
approximation rather than factual retrieval. Date matching is also the field most
sensitive to the matching threshold: under exact matching ($\theta=1.0$), F1
falls to 0.150 for birth dates and 0.134 for death dates, because at
$\theta=0.6$ dates that agree in year and month but differ in day are accepted
(Supplementary Table~\ref{tab:s_wiki}).

The \emph{influenced\_by} field highlights the model’s weakness on relational knowledge. With F1 = 0.103 and a hallucination rate of 89.6\,\%, the generated influence networks are largely inconsistent with the documented record. This illustrates structural distortion in relational biographical fields, consistent with the mechanism described in Section~\ref{sec:structural}.

\paragraph{Representative source--output comparisons.}
Two cached records make the field-level comparison concrete. For Kuno Fischer,
the Wikidata reference record gives the birth date \texttt{1824-07-23}, whereas
the cached reconstruction gives \texttt{1824-04-22}; the values do not satisfy
the local date-matching rule. The reference \emph{influenced\_by} field is
empty, whereas the reconstruction supplies \emph{Hegel} and \emph{Kant}; both
are local reference nonmatches. Thus, relative to this reference record, the
output substitutes a birth-date attribute value and adds two relation targets.

For Carl Stumpf, the cached birth date matches exactly
(\texttt{1848-04-21}), but the reference \emph{influenced\_by} value is
\emph{Bernard Bolzano} whereas the reconstruction gives \emph{Franz Brentano}.
The generated relation value does not satisfy the local comparison against the
reference value, so it replaces the recorded relation target in the
field--value representation. These examples are source-reference mismatches
under the recovered local rule. The item-level Wikidata matching record has been
regenerated from the cached reference and generated records with the original
matching functions; it reproduces the cached TP, FP, FN, and TN totals of every
field exactly and is included in the reproducibility package. The DBpedia split
of the false positives, by contrast, rests on the aggregate counts stored in the
evaluation cache, because the item-level DBpedia responses were not preserved.
The examples therefore do not by themselves establish whether a particular
unmatched generated value is verified extra knowledge, factually false, or a
limitation of the specified reference record.

\begin{figure}[H]
\centering
\includegraphics[width=1.00\textwidth]{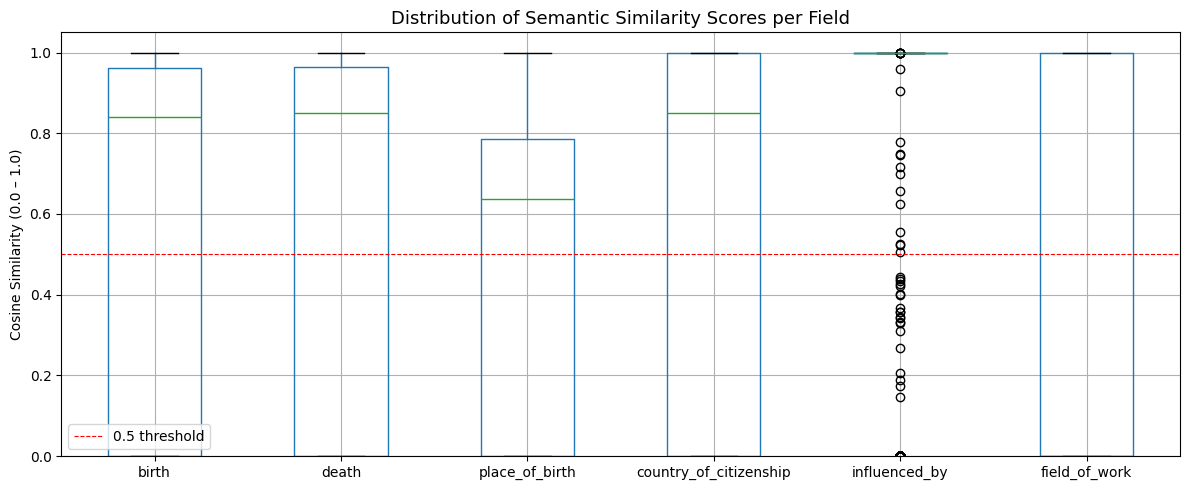}

\smallskip
\caption{Semantic similarity scores (cosine similarity using
\texttt{all-MiniLM-L6-v2}) for the philosopher benchmark. Even for fields where
the model achieves moderate F1 scores, the semantic similarity distributions are
wide, indicating high variance in the quality of LLM output across
philosophers. The wide distributions for \emph{influenced\_by} and
\emph{field\_of\_work} reflect the model's inconsistent handling of relational
and categorical fields.}
\label{fig:phil_semantic}
\end{figure}

\subsection{Benchmark 3: Bibliographic Knowledge --- Citations Retrieved from Dimensions.ai}
\label{sec:bibliographic_results}

The third benchmark evaluates the LLM's capacity to reconstruct bibliographic
data for 50 COVID-19 publications and the 638 records that cite them, sourced from the
Dimensions.ai API \citep{Dimensions2026} via the \texttt{dimcli} Python library.
For each one of these publications, the ten fields evaluated are: publication authors, title, year, date, DOI, type, concepts, research areas, times cited (i.e., the number of times the publication has been cited by other publications), and the list of citing publications (the \texttt{cited\_by} field, i.e.\ records that cite the publication, not the publication's own references).

\subsubsection{Citation Reconstruction Results}

For each publication in this reference bibliographic dataset, we construct a standardized reference string concatenating the following three fields: authors, title, and year. That string is then sent to the LLM with a strict instruction to return a JSON object containing seven specific bibliographic fields (date, DOI, type, concepts, research areas, times cited, and citations).
After the model’s response is parsed, cleaned, type-normalized, and converted into structured fields, we obtain an LLM-generated bibliographic record that can be directly aligned with and compared to the reference dataset.

The central quantitative finding of the bibliographic benchmark is the absence of recovered citation records. The reference snapshot
contains 638 citing records across the 50 publications (between 10 and 19 per
publication; 636 distinct records). The prompt asked for up to 20 citing
records per publication, so the full reference set was within reach of every
response. The LLM returned only 53 citing records in total, a count ratio of
8.3\,\%, and provided any at all for only 14 of the 50 publications (28\,\%),
leaving the remaining 36 publications (72\,\%) without citations. None of the
53 generated records matches a reference citing record of the same publication,
either by DOI or by title under the Dimensions equivalence test; the highest
title Jaccard similarity of any generated record to any reference record of its
publication is 0.55. Matched citation recall is therefore 0 of 638.

\begin{table}[!ht]  
\centering
\caption{Citation recall statistics for the bibliographic benchmark. A generated
citing record is matched if it shares the DOI of, or meets the Dimensions title
test with, a reference citing record of the same publication.}
\label{tab:bib_citations}
\begin{tabular}{lc}
\toprule
\textbf{Metric} & \textbf{Value} \\
\midrule
Reference citing records (Dimensions)        & 638 \\
LLM-generated citing records                 & 53 \\
Count ratio (generated / reference)          & 8.3\,\% \\
Generated records matched to a reference record & 0 \\
Matched recall                               & 0\,\% \\
Omission rate (unmatched reference records)  & 100\,\% \\
Papers with $>0$ LLM citations               & 14 / 50 (28\,\%) \\
\bottomrule
\end{tabular}
\end{table}

Because the requested number of records exceeded every reference list, the result is unlikely to be an artefact of the length instruction. Within this benchmark, the citation records were not reliably recovered through the tested parametric-generation procedure. As argued in Section~\ref{sec:structural}, this is reported as bibliographic structural divergence: the model does not reliably reproduce the reference citation records.

\paragraph{Representative citation-network comparisons.}
Two cached records make the aggregate result concrete. For \emph{The
contribution of the private healthcare sector during the COVID-19 pandemic: the
experience of the Lombardy Region in Northern Italy.}, the reference snapshot
contains 19 citing records, including \emph{The energy burden of care: Insights
from a regional survey in Italian hospitals.}
(\texttt{10.1177/09514848261416445}), whereas the cached LLM reconstruction
returns no citations. Relative to the experimental reference snapshot, this
omits all 19 observed citation edges for that record.

For \emph{Small fibre neuropathy frequently underlies the painful long-COVID
syndrome}, the reference snapshot contains 18 citations, including
\emph{Non-invasive electrodiagnostic testing for small fibre neuropathy in long
COVID-19} (\texttt{10.1136/bmjno-2025-001399}). The cached reconstruction
returns five citations, including \emph{Neuropathic manifestations in
post-COVID-19 patients: A systematic review}
(\texttt{10.1016/j.jns.2024.120123}). The retained evaluation records a
mean maximum title similarity of 0.757 across the five generated citations, a
year-match rate of 1.0, and a DOI-validity rate of 0.0. This is a
source--output citation-set and identifier mismatch under the cached comparison;
it does not establish that every generated title is fictitious or that each
unresolved DOI is an unresolved identifier.

\subsubsection{Field-Level Results}

\begin{table}[!ht] 
\centering
\caption{Classical evaluation results for the Dimensions bibliographic benchmark
($n=50$ publications). All scores are pooled over publications; the F1 interval
is a 95\% bootstrap interval over publications. No hallucination rate is
reported, because the Dimensions benchmark has no second reference source
against which unmatched values could be checked.}
\label{tab:bib_classical}
\begin{tabular}{lrrrccccc}
\toprule
\textbf{Field} & \textbf{TP} & \textbf{FP} & \textbf{FN} & \textbf{Precision} & \textbf{Recall} & \textbf{F1} & \textbf{95\% CI (F1)} & \textbf{Accuracy} \\
\midrule
date             & 15 &  35 &  35 & 0.300 & 0.300 & 0.300 & [0.180, 0.420] & 0.176 \\
doi              &  0 &  12 &  50 & 0.000 & 0.000 & 0.000 & [0.000, 0.000] & 0.000 \\
type             & 35 &  15 &  15 & 0.700 & 0.700 & 0.700 & [0.580, 0.820] & 0.538 \\
time\_cited      &  0 &  14 &  50 & 0.000 & 0.000 & 0.000 & [0.000, 0.000] & 0.000 \\
main\_concepts   & 55 & 190 & 187 & 0.224 & 0.227 & 0.226 & [0.173, 0.282] & 0.127 \\
research\_areas  &  2 & 243 & 127 & 0.008 & 0.016 & 0.011 & [0.000, 0.027] & 0.005 \\
\bottomrule
\end{tabular}
\end{table}

\begin{figure}[!ht] 
\centering
\includegraphics[width=\textwidth]{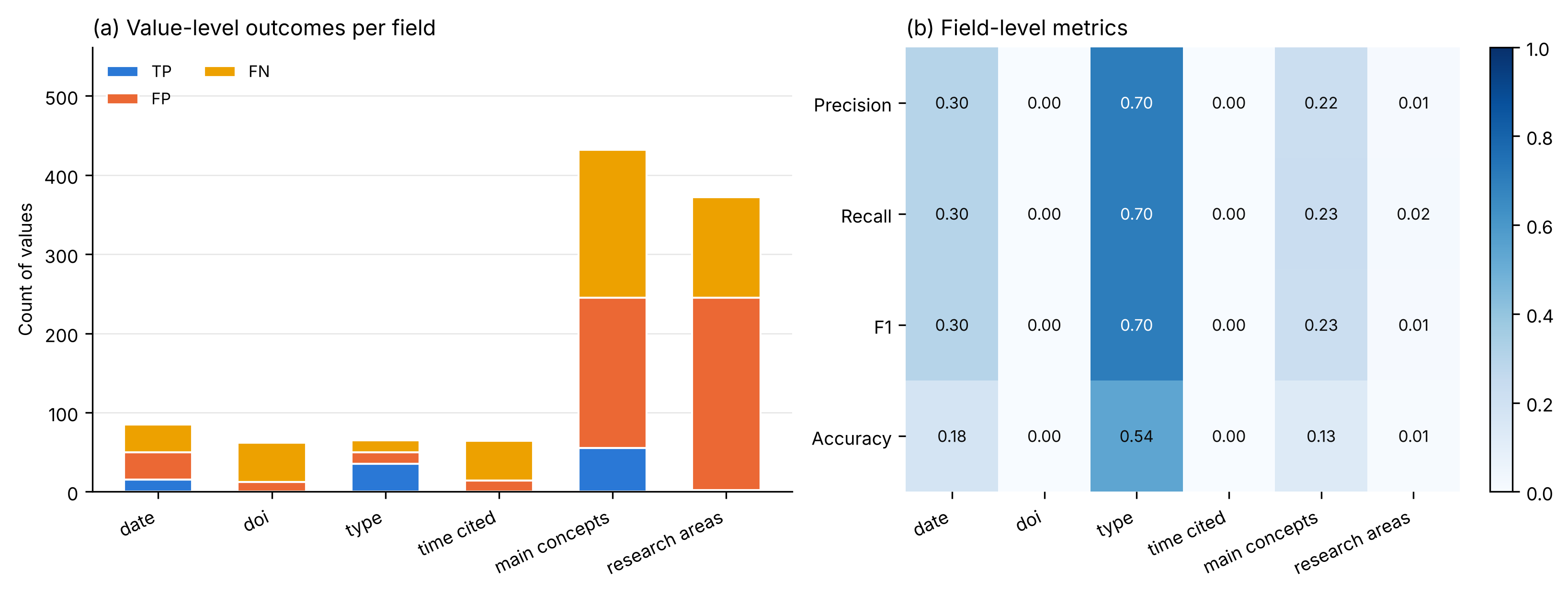}

\smallskip
\caption{Classical evaluation of LLM reconstruction of Dimensions bibliographic
data ($n=50$ publications). Panel (a) shows the pooled TP, FP, and FN counts
per field; panel (b) shows the pooled precision, recall, F1, and accuracy of
Table~\ref{tab:bib_classical}.}
\label{fig:bib_classical}
\end{figure}

The field-level results differ across the reported benchmark variables. In the
recovered 50-record source cache, publication type takes exactly three observed
normalised values: \texttt{article} (30 records), \texttt{book} (10), and
\texttt{chapter} (10). The corresponding cached LLM outputs also use these
three labels---\texttt{article} (45), \texttt{book} (3), and
\texttt{chapter} (2)---so the publication-type result (F1 = 0.70) is evaluated
within this sample-specific three-label set. This observed set is not claimed to
be a complete Dimensions taxonomy. The retained materials do not establish a
finite controlled range for date, DOI, times cited, main concepts, or research
areas; their lower scores are therefore reported without attributing them to an
unverified vocabulary size, hierarchy, or inherent inferability from title and
year.

The DOI failure is particularly instructive. A DOI is a unique, persistent
identifier that cannot be guessed or inferred; it must be looked up in a
registry \citep{Paskin2010}. The model left the DOI empty for 38 of the 50
publications and supplied one for 12; none of the 12 equals the reference DOI.
At the time of the run, 8 of the 12 did not resolve at \texttt{doi.org} and 4
resolved to other records. The zero F1 score for DOIs is consistent with the
model generating plausible-looking identifiers rather than retrieving them.
The date result depends strongly on the matching threshold. Both dates are
stored as timestamps, and the model usually returns 1 January of the
publication year; at $\theta=0.6$, token overlap therefore also accepts pairs
that share the year and the day ``01'' but differ in month. At $\theta=0.4$,
any pair with the same year matches and F1 rises to 0.96, whereas exact
matching gives 0.10 (Supplementary Table~\ref{tab:s_dim}). The model thus usually gives the
correct year but rarely the correct date, and the date F1 of 0.300 in
Table~\ref{tab:bib_classical} should be read as an upper bound on date-level
agreement.

\begin{figure}[H]
\centering
\includegraphics[width=1.00\textwidth]{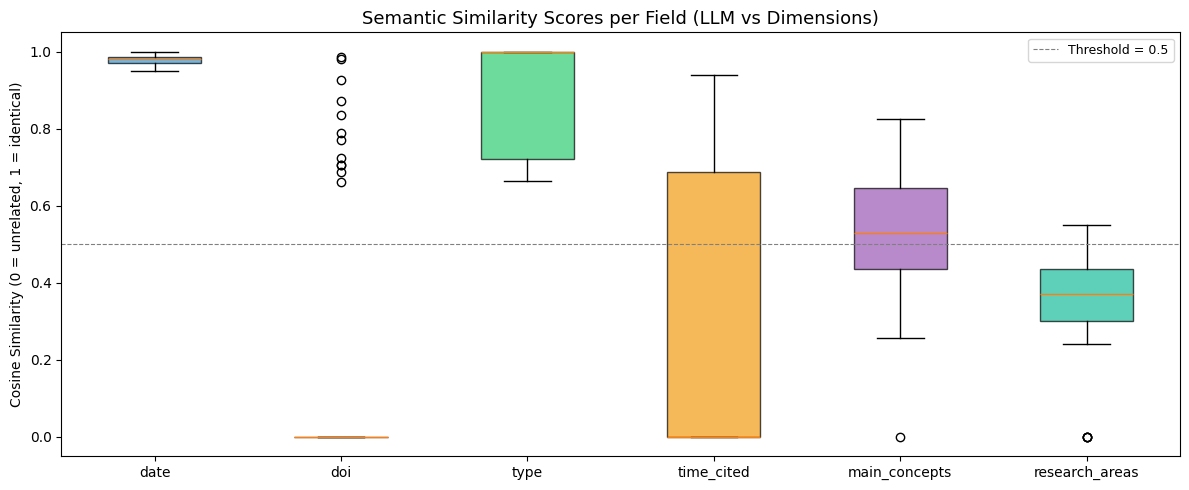}

\smallskip
\caption{Distribution of cosine-similarity scores by field for LLM-generated and reference Dimensions values, computed with \texttt{all-MiniLM-L6-v2}. Scores range from 0 (unrelated) to 1 (identical). The boxplots show the distributions for date, DOI, type, \texttt{time\_cited}, \texttt{main\_concepts}, and \texttt{research\_areas}; horizontal median lines, boxes, whiskers, and circular outliers follow the standard boxplot convention. The dashed horizontal line marks the similarity threshold of 0.5.}
\label{fig:bib_semantic}
\end{figure}

\subsection{Cross--Domain Structural Pattern}

Across the three benchmarks, a comparable pattern emerges under the stated evaluation procedures. In the lexical domain, the generated entries keep the thesaurus format prescribed by the prompt while replacing much of the historical vocabulary with semantically related contemporary terms. In the biographical domain, performance is relatively stronger for categorical fields with constrained value spaces and substantially weaker for relational fields requiring historically grounded network knowledge. In the bibliographic domain, identifier-based and citation-network fields exhibit extremely low recall, indicating non-retrievability of database-structured relations.

Taken together, these results suggest that parametric language models generate semantically plausible approximations without reliably reconstructing externally verifiable relational structures. Classical field-level metrics capture local mismatch, but graph-based diagnostics reveal structural divergence that is not visible through token-level evaluation alone. This cross-domain consistency supports the structural hallucination hypothesis advanced in Section~\ref{sec:structural}.

\medskip
\section{Discussion}
\label{sec:discussion}

\begin{table}[H]
\centering
\caption{Cross-domain structural divergence summary across lexical, biographical, and bibliographic benchmarks.}
\label{tab:structural_divergence_summary}
\small
\begin{tabular}{>{\raggedright\arraybackslash}p{0.17\textwidth}>{\raggedright\arraybackslash}p{0.43\textwidth}>{\raggedright\arraybackslash}p{0.08\textwidth}>{\raggedright\arraybackslash}p{0.2\textwidth}}
\toprule
\textbf{Domain} & \textbf{Source mismatch / hallucination rate} & \textbf{Node $J$} & \textbf{Citation omission} \\
\midrule
Lexical\newline (Roget) & 71.4\,\% source-nonmatched nodes; 0 of 262 obsolete terms recovered & 0.088 & -- \\
Biographical\newline (Wikidata) & 40.4--89.6\,\% of generated values per field absent from Wikidata and DBpedia & -- & -- \\
Bibliographic\newline (Dimensions) & 0 of 12 generated DOIs correct & -- & 100\,\% (0 of 638 matched) \\
\bottomrule
\end{tabular}
\end{table}

Table~\ref{tab:structural_divergence_summary} summarizes the divergences observed under the study's matching and evaluation procedures in three selected benchmark settings. The Roget case should be interpreted as a historical lexical stress test: its differences may reflect both non-recovery of the 1911 source structure and historical change in terminology. The Wikidata and Dimensions analyses provide separate contemporary tests involving biographical and bibliographic data. Across these cases, \texttt{gpt-4.1-mini} outputs show low alignment with the respective benchmark-defined relational structures, expressed as node or attribute mismatches and citation omissions. This convergence supports the usefulness of structural hallucination as a descriptive framework for the mismatches observed here; it does not establish a universal property of all knowledge organization systems, model architectures, or task settings.

Across the three benchmarks, the evaluated language model produced outputs that were often surface-plausible but did not reliably align with the benchmark-defined relational structures. In the historical Roget stress test, this mismatch must be interpreted in light of both non-recovery of the 1911 source structure and historical lexical change. Importantly, the biographical (Wikidata) and bibliographic (Dimensions) benchmarks provide independent contemporary evaluations, indicating that structural divergence is not confined to the historical lexical case. Taken together, these results demonstrate the value of examining structural divergence alongside local accuracy within this evaluation design. They support structural hallucination as a useful description of the observed failures, but do not establish that such divergence is universal across parametric models, knowledge organization systems, or task settings, or that it would persist under retrieval-augmented generation.

The empirical findings have three methodological implications.

First, LLM outputs should not be treated as substitutes for structured
database retrieval when precise bibliographic or factual reconstruction is
required. The observed source-mismatch and omission rates indicate that direct
verification against authoritative sources remains necessary.

Second, semantic similarity does not guarantee structural fidelity. Moderate
semantic alignment in the Roget benchmark coexists with low node-set overlap
($J=0.088$), even though the generated terms are demonstrably specific to the
Head they were requested for. This confirms the analytical distinction between local plausibility
and global relational alignment introduced earlier.

Third, network-analytic comparison provides diagnostic information not
captured by classical precision–recall metrics. Measures such as node-set
Jaccard similarity, centrality rank correlation, and modularity comparison
expose structural divergence that would remain invisible under sentence-level
evaluation alone.

\subsection{Applications: Assessment, Manuscript Writing, and Syllabus Design}
\label{sec:applications}

The same comparison principle can support source-aware assessment, literature-review checking, and AI-assisted metadata or syllabus design. In each case, the appropriate comparison source, scope, version, and matching rules must be explicit. The framework is intended as a verification method, not as a substitute for domain judgment or a claim that one reference source exhausts valid knowledge. In assessment, a student's or a model's concept map can be compared with a course reference graph using the node and edge overlap measures of Table~\ref{tab:kg_summary}; in manuscript writing, a draft's reference list can be checked against a bibliographic database in the manner of Table~\ref{tab:bib_citations}; and in syllabus design, generated reading lists and topic sequences can be compared with a curated curriculum graph. In each setting, the per-record matching tables produced by the pipeline indicate which elements need human review.

\subsection{Limitations and Scope Conditions}
\label{sec:limitations}

Several limitations bound the interpretation of these findings:

\begin{itemize}
    \item \textbf{Reference bias and domain dependence.} The reference graphs function as documented comparison baselines rather than as absolute truth. They embody source-specific scope, provenance, curatorial decisions, versioning, and possible omissions or biases. In domains without a canonical graph, researchers must construct a transparent expert reference or use consensus-based comparison. The present results may differ for other domains, model architectures, retrieval-augmented systems, or task formulations.
    \item \textbf{Historical bias.} The Roget analysis is a historical lexical stress test based on the 1911 edition. Its observed mismatch reflects both failure to recover the source structure and historical change in terminology. It does not provide independent evidence about contemporary knowledge organization systems.
    \item \textbf{Matching sensitivity.} The strict operational matching used here may penalize semantic variation. The fuzzy matching threshold may undercount partial lexical correspondence, and the DBpedia oracle is not exhaustive. All results are conditional on the stated matching thresholds; alternative matching rules or semantic embedding thresholds would yield different absolute mismatch rates. The sensitivity analyses in Supplementary Appendix S1 show that, for thresholds between 0.4 and 1.0, the F1 scores of all non-date Wikidata fields vary by at most 0.005, whereas date fields in both Wikidata and Dimensions depend materially on the threshold, as do the Dimensions concept fields (F1 for main concepts between 0.18 and 0.30); the DOI and times-cited results are zero at every threshold, and the citation result remains zero down to $\theta=0.6$; the one to three title matches that appear at $\theta=0.5$ or $0.4$ pair different papers on similar topics.
    \item \textbf{Prompt constraints and single model.} The Roget prompt limited each generated list to 10--20 items, which bounds attainable recall; precision and the node-overlap results are less affected. All benchmarks use one model (\texttt{gpt-4.1-mini}) at temperature 0 with one response per record, so the results do not describe run-to-run variability or other models.
    \item \textbf{Item-level DBpedia evidence.} The hallucination rates in Table~\ref{tab:phil_classical} rest on the aggregate DBpedia counts preserved in the evaluation cache; the item-level DBpedia responses were not preserved, although the item-level Wikidata matching record has been regenerated and reconciles exactly with the cached totals.
    \item \textbf{Contemporary KOS coverage.} Wikidata and Dimensions provide contemporary knowledge-graph and bibliographic-graph cases, but they do not substitute for a contemporary KOS-specific replication. A benchmark using a current controlled vocabulary, thesaurus, or ontology is needed to isolate structural reconstruction from historical lexical change.
\end{itemize}

These limitations restrict the present claims to the evaluated model, benchmarks, and evaluation design while identifying priorities for replication and sensitivity analysis.

\medskip
\section{Conclusion}
\label{sec:conclusion}

This paper advanced a conceptual account of structural hallucination—distortion at the level of relational organization—and a set of empirical benchmarks to detect it. We argued that sentence-level verification is inadequate for domains where knowledge is organized relationally and introduced a reproducible validation architecture grounded in knowledge graph extraction, network-analytic diagnostics, and citation integrity verification.

Within the three benchmark settings examined here, \texttt{gpt-4.1-mini} outputs showed substantial divergence from the documented comparison structures: limited node-set overlap in the historical Roget stress test, high biographical field mismatch in Wikidata, and citation omission in Dimensions. The Roget result must be interpreted in light of both non-recovery of the 1911 source structure and historical change in terminology. These findings show that semantic plausibility need not guarantee structural fidelity under the study's extraction and matching procedures, and that graph-level evaluation can complement local accuracy metrics when a documented reference structure is available.

Beyond research evaluation, the framework has direct implications for digital library governance. As libraries experiment with AI-assisted metadata generation, structural hallucination risks propagating into catalogues and discovery layers. The hallucination stress test can be operationalised as a pre-ingestion quality-control layer: curators could extract a knowledge graph from generated records, compare it against authoritative reference graphs, and compute source-mismatch and omission rates. Records exceeding structural deviation thresholds would trigger human review, functioning analogously to existing authority control procedures \citep{Borgman1999, Tennis2012}. Such integration would transform LLM oversight from reactive correction to proactive infrastructural design.
The framework and benchmarks presented here provide a methodological basis for evaluating structural alignment in LLM-generated scholarly representations. Future research may extend the approach to additional domains, incorporate retrieval-augmented architectures, and explore the interaction between parametric generation and external knowledge graphs. The broader challenge is not whether LLMs can produce fluent text, but how their outputs can be assessed for structural compatibility with established knowledge systems.

\section*{Generative AI Use Statement}

Uses and Roles of Generative AI: The manuscript evaluates structured outputs produced by gpt-4.1-mini in the reported benchmark experiments. Claude (Anthropic) was also used during preparation of the revision for drafting, formatting, and code-documentation assistance. The author reviewed and approved all AI-assisted material and remain fully responsible for the manuscript's content.

\section*{Data Availability Statement}

The code-only reproducibility package for this study is maintained at
\url{https://github.com/mboudour/Network_Based_LLMs_Hallucination_Eval}
(private repository; editorial and reviewer access is available from the
authors on request). It contains the analysis code, fixed benchmark prompts,
model-configuration information, normalization and matching routines,
figure-generation scripts, and documentation needed to inspect the study
workflow.

For this revision, the package also contains, in its folder \texttt{release\_r2}, which is provided to the editor and reviewers as a supplementary file, the recomputation scripts that
produce every table and figure from the cached records, the item-level matching
records of the Roget and Wikidata benchmarks (one row per record and field, with
the TP, FP, FN, and TN assignment), and the Roget and Wikidata reference and
generated records on which they are computed. The Wikidata item-level record
reproduces the cached field totals exactly; the DBpedia split of its false
positives is available only as the aggregate counts stored in the evaluation
cache, because the item-level DBpedia responses were not preserved.

The repository otherwise omits processed Dimensions data, cached provider
records, query logs, and the Dimensions item-level matching records. The principal restriction concerns the Dimensions benchmark:
Dimensions bibliographic and citation records are governed by provider access
and redistribution terms. The cached source records, query logs, and
record-level matching materials are therefore not publicly deposited where
they reproduce, or could permit reconstruction of, restricted provider
content. The corresponding raw LLM outputs are retained with that controlled
record-level evaluation archive rather than released as a separate public
dataset. These restrictions concern source licensing and provenance; they do
not prevent methodological inspection of the code, prompts, configurations,
normalization rules, and figure-generation workflow.

The Roget source text is available from Project Gutenberg
(\url{https://www.gutenberg.org/ebooks/10681}), and Wikidata reference records
can be re-queried through the Wikidata Query Service
(\url{https://query.wikidata.org}). The rendered figures are included in the
manuscript and can be regenerated from the deposited code by users with the
relevant source access. Where permitted by the applicable source terms, the
controlled archive may be made available to the editor and reviewers on
request.

\bibliographystyle{apalike}
\bibliography{references,references_additions}

\clearpage
\appendix
\renewcommand{\thetable}{S\arabic{table}}
\renewcommand{\thefigure}{S\arabic{figure}}
\setcounter{table}{0}
\setcounter{figure}{0}
\section*{Supplementary Appendix S1}
\addcontentsline{toc}{section}{Supplementary Appendix S1}

\subsection*{S1.1 Notation key}

\begin{table}[H]
\centering
\caption{Notation used in the matching rule of Section~\ref{sec:matching}.}
\label{tab:s_notation}
\small
\begin{tabular}{p{0.2\textwidth}p{0.72\textwidth}}
\toprule
\textbf{Symbol} & \textbf{Meaning} \\
\midrule
$b$ & Benchmark: Roget, Wiki (Wikidata philosophers), or Dim (Dimensions publications). \\
$u$, $v$ & A generated value and a reference value (a term, attribute value, or citing record). \\
$r(\cdot)$ & The record containing a value: a Roget Head, a philosopher, or a publication. \\
$A_b(r(u),r(v))$ & 1 if the two records share the benchmark key (normalised Head name; exact philosopher label; publication title), else 0. \\
$\rho(\cdot)$ & The field or relation role carrying the value (e.g.\ \texttt{noun\_list}, \texttt{birth}, \texttt{cited\_by}). \\
$\nu(\cdot)$ & Lower-casing and trimming. \\
$T(\cdot)$ & Set of tokens of $\nu(\cdot)$ after splitting at spaces, hyphens, and commas. \\
$\sqsubseteq$ & Substring containment. \\
$E_b(u,v)$ & Benchmark-specific value-equivalence test (Section~\ref{sec:matching}). \\
$M_b(u,v)$ & Matching rule: $A_b=1 \land \rho(u)=\rho(v) \land E_b=1$. \\
$\theta$ & Token-overlap threshold in $E_{\mathrm{Wiki}}$ and $E_{\mathrm{Dim}}$ (0.6 in the main text). \\
TP, FP, FN, TN & Pooled counts per field; FP is split into $\mathrm{FP}_{\mathrm{extra}}$ and $\mathrm{FP}_{\mathrm{hall}}$ in the Wikidata benchmark. \\
\bottomrule
\end{tabular}
\end{table}

\subsection*{S1.2 Fixed prompts}

All calls used \texttt{gpt-4.1-mini} at temperature 0. The Roget call used a 1,024-token limit and JSON response mode; the user message was ``Head name: \emph{name}''.

\noindent\textbf{Roget (system prompt).}
\begin{quote}\small\ttfamily\raggedright You are an expert in historical English lexicography and the 1911 edition of Roget's Thesaurus. Given a Head name, reconstruct the full thesaurus entry. Return a JSON object with exactly these keys: class, section, head\_name, noun\_list, verb\_list, adjective\_list, adverb\_list, cross\_references. Keep each list concise (10-20 items max) to avoid truncation.\end{quote}

\noindent\textbf{Wikidata philosophers (system prompt; 20 names per call).}
\begin{quote}\small\ttfamily\raggedright You are a philosophy historian. Given a JSON array of philosopher names, return a JSON array of objects -- one object per philosopher, in the same order. Each object must have exactly these keys: "philosopher" -- the name exactly as given; "birth" -- ISO date YYYY-MM-DD, or empty string if unknown; "death" -- ISO date YYYY-MM-DD, or empty string if unknown; "place\_of\_birth" -- city or region, or empty string if unknown; "country\_of\_citizenship" -- country name, or empty string if unknown; "influenced\_by" -- comma-separated names, or empty string if unknown; "field\_of\_work" -- comma-separated fields, or empty string if unknown. Respond ONLY with the JSON array. No markdown, no code fences, no extra text.\end{quote}
The user message was ``Philosophers: \emph{JSON array of names}. Return a JSON array only.''

\noindent\textbf{Dimensions publications (system prompt).}
\begin{quote}\small\ttfamily\raggedright You are a bibliographic expert. Given a paper reference, return a JSON object with exactly these keys: "date" -- release date as YYYY-MM-DD string; "doi" -- DOI string (without https://doi.org/ prefix); "type" -- publication type (article / book / chapter); "main\_concepts" -- list of top 5 concept strings; "research\_areas" -- list of research area strings (no leading numbers); "time\_cited" -- integer citation count; "cited\_by" -- list of up to 20 dicts, each with keys "citation\_title", "citation\_year", "citation\_doi". Return ONLY the JSON object. No markdown, no code fences, no extra text.\end{quote}
The user message was the reference string ``\emph{authors}, `\emph{title}' (\emph{year})''.

\subsection*{S1.3 Pseudocode}

\begin{quote}\small
\textbf{Input:} reference records $\mathcal{R}$, generated records $\mathcal{G}$, fields $F$, equivalence test $E_b$.\\
1.\ Align: for each $g\in\mathcal{G}$, find $r\in\mathcal{R}$ with $A_b(r,g)=1$; discard unaligned records.\\
2.\ For each aligned pair $(r,g)$ and field $f\in F$: normalise the values of $r$ and $g$ in field $f$ into sets $V_r$ and $V_g$.\\
3.\ \hspace*{1em}If $V_r=V_g=\emptyset$: TN $\mathrel{+}=1$.\\
4.\ \hspace*{1em}For each $u\in V_g$: if $\exists v\in V_r$ with $E_b(u,v)=1$, TP $\mathrel{+}=1$; otherwise FP $\mathrel{+}=1$ and, in Wikidata, query DBpedia for $(r,f)$: if $u$ matches a DBpedia value, $\mathrm{FP}_{\mathrm{extra}}\mathrel{+}=1$, else $\mathrm{FP}_{\mathrm{hall}}\mathrel{+}=1$.\\
5.\ \hspace*{1em}For each $v\in V_r$ with no $u\in V_g$ such that $E_b(u,v)=1$: FN $\mathrel{+}=1$.\\
6.\ Sum counts over records per field; compute $P$, $R$, $F_1$, accuracy, and (Wikidata) $\mathrm{FP}_{\mathrm{hall}}/(\mathrm{TP}+\mathrm{FP})$.\\
7.\ Bootstrap: resample aligned records with replacement 2,000 times and recompute $F_1$; report the 2.5th and 97.5th percentiles.\\
8.\ Graphs (Roget): add one Head node per record, one edge from the Head to each normalised term, and one edge per cross-reference; compare node and edge sets by Jaccard similarity.
\end{quote}
For scalar fields, step 4 and step 5 are applied with $|V_r|,|V_g|\le 1$, so a non-matching pair contributes one FP and one FN.

\subsection*{S1.4 Implementation checklist}

\begin{enumerate}\small
\item Record the reference source, version, and retrieval date (Gutenberg \#10681; Wikidata SPARQL snapshot of philosophers born 1800--1850; Dimensions snapshot of 50 COVID-19 publications).
\item Fix the sample (Roget: 30 Heads, Python seed 42) and the model settings (\texttt{gpt-4.1-mini}, temperature 0).
\item Cache every reference and generated record before evaluation.
\item Apply the normalisation and the equivalence test $E_b$ of Section~\ref{sec:matching}; save the item-level TP/FP/FN/TN assignment.
\item Reconcile item-level totals with the field-level table before reporting.
\item Report pooled scores, the denominators used, bootstrap intervals, and threshold sensitivity.
\item Build graphs from the same normalised values as the field-level evaluation.
\end{enumerate}

\subsection*{S1.5 Sensitivity analyses}

\begin{table}[H]
\centering
\caption{Roget: pooled F1 under three normalisations. ``List segments'' compares whole semicolon-separated segments (the protocol of the first version of this analysis); ``single terms'' is the main-text protocol; ``single terms, cleaned'' additionally removes obsolescence daggers and trailing punctuation.}
\label{tab:s_norm}
\small
\begin{tabular}{lccc}
\toprule
\textbf{Field} & \textbf{List segments} & \textbf{Single terms} & \textbf{Single terms, cleaned} \\
\midrule
noun\_list & 0.022 & 0.154 & 0.164 \\
verb\_list & 0.020 & 0.121 & 0.126 \\
adjective\_list & 0.018 & 0.154 & 0.166 \\
adverb\_list & 0.000 & 0.071 & 0.076 \\
cross\_references & 0.044 & 0.089 & 0.089 \\
\bottomrule
\end{tabular}
\end{table}

\begin{table}[H]
\centering
\caption{Wikidata: pooled F1 by field for token-overlap thresholds $\theta$ in $E_{\mathrm{Wiki}}$ (main text: $\theta=0.6$). Containment matches are accepted at every threshold.}
\label{tab:s_wiki}
\small
\begin{tabular}{lccccccc}
\toprule
\textbf{Field} & 0.4 & 0.5 & 0.6 & 0.7 & 0.8 & 0.9 & 1.0 \\
\midrule
birth & 0.306 & 0.306 & 0.214 & 0.150 & 0.150 & 0.150 & 0.150 \\
death & 0.296 & 0.296 & 0.217 & 0.134 & 0.134 & 0.134 & 0.134 \\
place\_of\_birth & 0.136 & 0.136 & 0.136 & 0.135 & 0.135 & 0.135 & 0.135 \\
country\_of\_citizenship & 0.588 & 0.588 & 0.586 & 0.586 & 0.586 & 0.586 & 0.586 \\
influenced\_by & 0.103 & 0.103 & 0.103 & 0.100 & 0.098 & 0.098 & 0.098 \\
field\_of\_work & 0.240 & 0.240 & 0.236 & 0.236 & 0.236 & 0.236 & 0.236 \\
\bottomrule
\end{tabular}
\end{table}

\begin{table}[H]
\centering
\caption{Dimensions: pooled F1 by field for Jaccard thresholds $\theta$ in $E_{\mathrm{Dim}}$ (main text: $\theta=0.6$).}
\label{tab:s_dim}
\small
\begin{tabular}{lccccccc}
\toprule
\textbf{Field} & 0.4 & 0.5 & 0.6 & 0.7 & 0.8 & 0.9 & 1.0 \\
\midrule
date & 0.960 & 0.340 & 0.300 & 0.300 & 0.100 & 0.100 & 0.100 \\
doi & 0.000 & 0.000 & 0.000 & 0.000 & 0.000 & 0.000 & 0.000 \\
type & 0.700 & 0.700 & 0.700 & 0.700 & 0.700 & 0.700 & 0.700 \\
time\_cited & 0.000 & 0.000 & 0.000 & 0.000 & 0.000 & 0.000 & 0.000 \\
main\_concepts & 0.299 & 0.277 & 0.226 & 0.186 & 0.182 & 0.182 & 0.182 \\
research\_areas & 0.037 & 0.027 & 0.011 & 0.000 & 0.000 & 0.000 & 0.000 \\
\bottomrule
\end{tabular}
\end{table}

\subsection*{S1.6 Head-level overlap}

\begin{table}[H]
\centering
\caption{Roget: Jaccard similarity between the generated and the reference term sets of each sampled Head (four term fields combined), in ascending order.}
\label{tab:s_heads}
\small
\begin{tabular}{lclc}
\toprule
\textbf{Head} & $J$ & \textbf{Head} & $J$ \\
\midrule
dextrality & 0.000 & ingratitude & 0.083 \\
pitfall & 0.015 & adversity & 0.086 \\
greatness & 0.019 & obstinacy & 0.089 \\
director & 0.037 & inequality & 0.091 \\
abrogation & 0.039 & language & 0.092 \\
narrowness. thinness & 0.043 & quadruplication & 0.097 \\
covering & 0.044 & chronometry & 0.097 \\
contention & 0.046 & unintelligibility & 0.101 \\
plainness & 0.047 & cessation & 0.112 \\
sequence & 0.051 & repetition & 0.116 \\
degree & 0.060 & superiority & 0.120 \\
convexity & 0.069 & good & 0.127 \\
duplication & 0.070 & permission & 0.134 \\
salubrity & 0.073 & brown & 0.189 \\
interposition & 0.076 & prohibition & 0.195 \\
\bottomrule
\end{tabular}
\end{table}

\subsection*{S1.7 Supplementary figure}

\begin{figure}[H]
\centering
\includegraphics[width=0.9\textwidth]{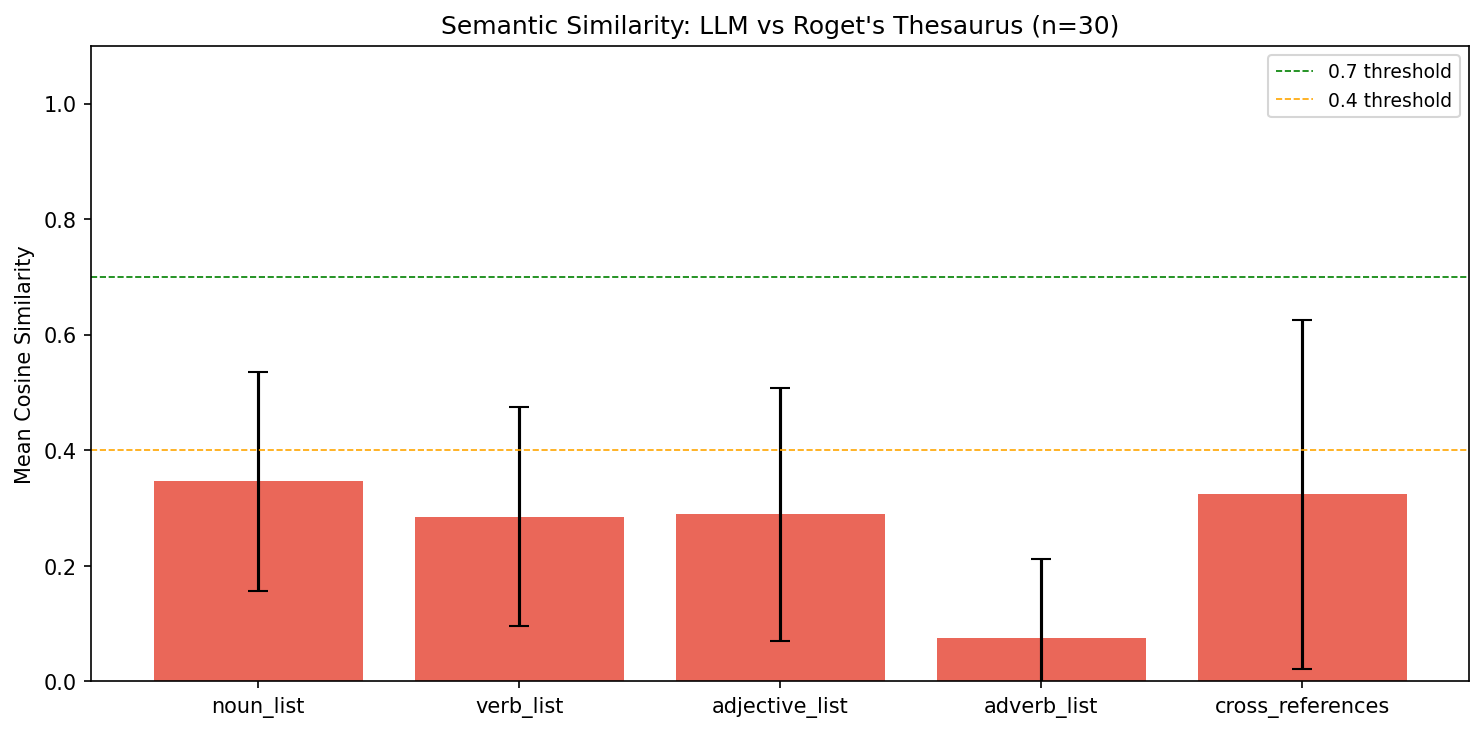}
\caption{Roget: mean cosine similarity (\texttt{all-MiniLM-L6-v2}) between generated and reference lists per field, with standard deviations ($n=30$ Heads). The similarity scores are descriptive and were not used to establish matches.}
\label{fig:s1}
\end{figure}

\end{document}